# Controllable Defect Driven Symmetry Change and Domain Structure Evolution in BiFeO$_3$ with Enhanced Tetragonality

Chao Chen,[†,‡] Changan Wang,[§] Xiangbin Cai,[∥] Chao Xu,[⊥] Caiwen Li,[†,‡] Jingtian Zhou,[#] Zhenlin Luo,[#] Zhen Fan,[†,‡] Minghui Qin,[†] Min Zeng,[†] Xubing Lu,[†] Xingsen Gao,*[,†] Ulrich Kentsch,[§] Ping Yang,[∇] Guofu Zhou,[‡,¶] Ning Wang,[∥] Ye Zhu,[⊥] Shengqiang Zhou,[§] Deyang Chen,*[,†,‡,¶] Jun-Ming Liu[●]

[†]Institute for Advanced Materials, South China Academy of Advanced Optoelectronics, South China Normal University, Guangzhou 510006, China

[‡]Guangdong Provincial Key Laboratory of Optical Information Materials and Technology, South China Academy of Advanced Optoelectronics, South China Normal University, Guangzhou 510006, China

[§]Helmholtz-Zentrum Dresden-Rossendorf, Institute of Ion Beam Physics and Materials Research, Bautzner Landstr. 400, 01328 Dresden, Germany

[∥]Department of Physics and Center for Quantum Materials, The Hong Kong University of Science and Technology, Clear Water Bay, Kowloon, Hong Kong, China

[⊥]Department of Applied Physics, The Hong Kong Polytechnic University, Hung Hom, Kowloon, Hong Kong, China

[#]National Synchrotron Radiation Laboratory, University of Science and Technology of China, Hefei, Anhui 230026, China

[∇]Singapore Synchrotron Light Source, National University of Singapore, 5 Research Link, 117603, Singapore

[¶]National Center for International Research on Green Optoelectronics, South China Normal University, Guangzhou 510006, China

[●] Laboratory of Solid State Microstructures and Innovation Center of Advanced Microstructures, Nanjing University, Nanjing 210093, China

## ABSTRACT

Defect engineering has been a powerful tool to enable the creation of exotic phases and the discovery of intriguing phenomena in ferroelectric oxides. However, accurate control the concentration of defects remains a big challenge. In this work, ion implantation, that can provide controllable point defects, allows us the ability to produce a controlled defect driven true super-tetragonal (T) phase with single-domain-state in ferroelectric $BiFeO_3$ thin films. This point-defect engineering is found to drive the phase transition from the as-grown mixed rhombohedral-like (R) and tetragonal-like ($M_C$) phase to true tetragonal (T) symmetry and induce the stripe multi-nanodomains to single domain state. By further increasing the injected dose of He ion, we demonstrate an enhanced tetragonality super-tetragonal (super-T) phase with the largest c/a ratio ~ 1.3 that has ever been experimentally achieved in $BiFeO_3$. A combination of morphology change and domain evolution further confirm that the mixed $R/M_C$ phase structure transforms to the single-domain-state true tetragonal phase. Moreover, the re-emergence of R phase and in-plane nanoscale multi-domains after heat treatment reveal the memory effect and reversible phase transition and domain evolution. Our findings demonstrate the reversible control of R-Mc-T-super T symmetry changes (leading to the creation of true T phase $BiFeO_3$ with enhanced tetragonality) and multidomain-single domain structure evolution through controllable defect engineering. This work also provides a pathway to generate large tetragonality (or c/a ratio) that could be extended to other ferroelectric material systems (such as $PbTiO_3$, $BaTiO_3$ and $HfO_2$) which might lead to strong polarization enhancement.

As a consequence of strong couplings and complex interplay between defects and the lattice, spin, charge and orbital degrees of freedom in complex oxides, defect engineering provides a plethora of splendid possibilities for creating a wealth of exotic phases and emergent physical phenomena.[1-7] In ferroelectric oxides, defect engineering is emerging as a powerful pathway to tailor structural transformation and manipulate material properties,[8-12] e.g., defect induced super-tetragonal $PbTiO_3$ and tetragonal-like $BiFeO_3$ thin films with giant tetragonality and enhanced polarization by the introduction of $PbO$[13] and $Bi_2O_3$ impurities,[14-16] respectively. The creation of novel domain states, the control of domain structure evolution, domain wall conduction and polarization switching as well as the enhancement of ferroelectric Curie temperature have also been realized through defect engineering.[17-22] However, these defects, introduced by changing growth conditions, cannot be precisely controlled. In recent years, the atomic-scale layer-by-layer growth techniques of oxide heterostructures have enabled the controllable introduction of planar defects—interfaces, leading to new modalities arising in superlattices, such as our recent discovery of polar vortices in $PbTiO_3/SrTiO_3$ superlattices[8] and domain configuration design in $BiFeO_3/La-BiFeO_3$ superlattices.[23] However, the post growth control of these planar defects is not allowed as they are imparted during the growth process. Therefore, ion implantation, the standard technique to introduce accurately controlled defects post growth in semiconductor devices for the adjustment of electronic properties,[24-26] is used in this work to study the controllable defect engineering of structure transformations and domain evolutions in ferroelectric $BiFeO_3$ thin films.

As one of the very few known room temperature multiferroic (ferroelectric and antiferromagnetic) materials,[27,28] $BiFeO_3$ exhibits rich phase diagrams include bulk rhombohedral, rhombohedral-like, tetragonal-like and orthorhombic phases that have been theoretically predicted by density functional calculations and phase field simulations,[29-31] and experimentally achieved through strain engineering (e.g., epitaxial strain and hydrostatic pressure)[29-39] and defect engineering (such as substitutional defects and impurity defects),[14,15,40-42] leading to a large number of intriguing properties.[29,43-51] Among those, the tetragonal-like $BiFeO_3$ driven by epitaxial strain with large c/a ratio (~ 1.23) is of great interest due to its giant spontaneous polarization up to 150 $\mu C/cm^2$.[42] Indeed, both theoretical calculations[52-56] and experimental studies[13,14,42,57-59] have demonstrated that the ferroelectric oxides with giant tetragonality (or c/a ratio) possess a large polarization as a consequence of their large dipolar moment. However, despite considerable effort and dramatic progress in the research of tetragonal-like (Mc) $BiFeO_3$, broad questions remain: (i) is it possible to stabilize the tetragonal-like phase $BiFeO_3$ in thick films while rhombohedral-like (R) phase generally emerges due to strain relaxation in the films thicker than ~30 nm,[29] (ii) is there a pathway to achieve the theoretically predicted true tetragonal (T) phase with single domain state instead of the experimentally observed tetragonal-like Mc phase with multi-nanodomain structure in pure $BiFeO_3$, (iii) could the c/a ratio (tetragonality) be further enhanced to obtain real super-tetragonal (super T) phase $BiFeO_3$ that would lead to a larger spontaneous polarization? These challenges to be addressed are the main focus of this study.

Ion implantation, as a controllable defect engineering route, has enabled the ability to continuously tailor the structure phase transition, electric transport, magnetic properties, as well as band gap in epitaxial oxide thin films by driving the out-of-plane lattice expansion while leaving the in-plane lattice epitaxially locked to the substrate.[60-64] Very recently, it has also been demonstrated that He ion implantation can be used to control ferroelectric switching behaviors and enhance electrical resistivity in ferroelectric thin films (such as $PbTiO_3$, $PbZr_{0.2}Ti_{0.8}O_3$ and $BiFeO_3$).[65-67] However, the use of He implantation to simultaneously control phase transition and domain structure evolution in $BiFeO_3$ thin films remain elusive.

In this work, we start from $BiFeO_3$ films with a mixed rhombohedral-like (R) and tetragonal-like (Mc) Phase. We have discovered that controlled point defects, introduced by He ion implantation, can be used to drive R-Mc-T-super T phase transition, enabling the formation of true super-tetragonal phase in 70-nm-thick $BiFeO_3$ films with enhanced tetragonality (c/a ratio ~ 1.3). Implanted He ions in the films enable not only the absence of R phase but the transformation from Mc structure to theoretical predicted true T symmetry.[52,53] Further increasing the dosages can continuously enhance the c-axis lattice parameter from 4.65 Å to 4.93 Å as well as the c/a ratio from 1.22 to 1.3 (the largest tetragonality has ever been experimentally obtained in $BiFeO_3$). It is also found that the in-plane nanoscale stripe domains (multi-domains) in the as-grown sample evolve to single domain state after He implantation, which further confirms the Mc to true T phase transition. Moreover, the annealing process has enabled the re-emergence of R phase and in-plane stripe nanodomains, demonstrating the reversible control of phase transition via He implantation.

We grew a series of epitaxial 70-nm-thick $BiFeO_3$ films on $LaAlO_3$ (001) substrates by pulsed laser deposition (PLD) at 680 °C in an $O_2$ pressure of 100 mTorr and cooled in a 1 atm oxygen atmosphere. After deposition, He ion implantation was carried out at 8 keV with a fluence from $5 \times 10^{14}$ to $5 \times 10^{15}$ ions/cm². A combination of X-ray diffraction (XRD), reciprocal space mapping (RSM) and transmission electron microscopy (TEM) was used for the structural characterization. Morphology changes and ferroelectric domain structures were measured using atomic force microscopy (AFM) and piezoresponse force microscopy (PFM).

The topography of the as-grown sample, as shown in Figure 1a, was imaged by AFM measurement, revealing the coexistence of striped dark contrast R phase and bright contrast $M_C$ phase matrix in the 70-nm-thick $BiFeO_3$ film, which is consistent with previous work.[29] The atomically flat terraces, one unit cell in height, confirm the high quality growth of the film, with a root mean square (RMS) roughness only of 0.2 nm. Interestingly, the striped R phase is greatly reduced after He implantation with the dose of $5 \times 10^{14}$ ions/cm$^2$, and completely disappears as the implanted dose up to $5 \times 10^{15}$ ions/cm$^2$, as presented in Figure 1b and 1c, respectively. We infer that there is a R-Mc phase transition induced *via* He implantation. To verify this hypothesis, X-ray reciprocal space maps (RSMs) are carried out. RSM (002) reflections in Figure 1d show the coexistence of R and T-like phases and three-fold and two-fold splits along (103) and (113) RSMs, as shown in Figure 1g and 1h, confirm that the tetragonal-like structure is Mc phase in the as-grown sample in agreement with previous work.[31,35,39] Surprisingly, the mixed R and Mc phases evolve to true tetragonal structure with enhanced c/a ratio after He implantation (Figure 1e, i and j). Although the nearly tetragonal phase was obtained by biaxial strain[68,69], this is the first report to achieve true T phase in high quality pure $BiFeO_3$ at room temperature, which has only been realized at high temperature or in chemical doping $BiFeO_3$.[31,70,71] Moreover, further enhancement of tetragonality (c/a ratio ~ 1.29) is obtained as the implanted He dose increases to $5 \times 10^{15}$ ions/cm$^2$. Here we refer this tetragonal phase with ultra-large c/a ratio to true super-tetragonal (super-T) phase, which is different from previous reported super-tetragonal like phase that is Mc phase in fact.[29,33,35,39,72,73] Typical X-ray θ–2θ scans show obvious peak shift with the increasing implanted dosages, as shown in Figure 1k, further confirming the Mc-T-super T symmetry changes driven by this controlled defect engineering route. Based on these XRD data (Figure 1d-k), the dependence of c/a ratio and c-axis with various He dosages is plotted in Figure 1l. It shows the continuous enhancement of c-axis lattice parameter from 4.65 Å to 4.93 Å as well as the c/a ratio from 1.22 to 1.29. Those are the largest values of c-axis and c/a ratio that have ever been experimentally achieved in $BiFeO_3$.

To understand the fine structure of these implanted films, atomic resolution annular dark-field scanning transmission electron microscopy (ADF-STEM) studies were carried out. Figure 2a shows a dark field cross-sectional STEM image of the implanted film with $2.5 \times 10^{15}$ He/cm$^2$, not only confirming the thickness (~ 70 nm) of $BiFeO_3$ and the existence of a single phase (no contrast differences as prior work[29, 71]), but also revealing the maintenance of high-quality epitaxy with sharp interfaces after He implantation. A high-resolution Z-contrast image of the $BiFeO_3$/$LaAlO_3$ interface is shown in in the bottom-right insets of Figure 1a, illustrating the atomic-scale epitaxy between the $BiFeO_3$ and $LaAlO_3$ layers. This type of image is sensitive to variations in atomic number and the intensity corresponds to increasing Z number. Thus, the elements in decreasing order of brightness are Bi (83) and Fe (26), while O (8) is too light to be visible, which is confirmed in Figure 1b using an ADF-STEM image and the corresponding EDS mappings. The selected area electron diffraction (SAED) patterns of implanted $BiFeO_3$ with He doses of $2.5 \times 10^{15}$ and $5 \times 10^{15}$ ions/cm$^2$ in Figure 2c and d, reveal the existence of single T phase (*i.e.*, absence of R phase) and the enhancement of c/a ratios to 1.27 and 1.30, respectively. Further atomic structure of these implanted films is shown in Figure 2e and f, presenting the strongly elongated out-of-plane lattice parameter driven by He implantation with the almost no changes of in-plane lattice parameter constrained by the substrate. Specifically, the c-axis changes from 4.65 Å to 4.93 Å and the c/a ratio is enhanced from 1.22 to 1.30, in contrast to the as-grown sample.

After successfully realizing the defect induced true super-tetragonal phase with giant tetragonality (c/a ~ 1.30) in $BiFeO_3$ films, we next turn to study the domain structure evolution driven by this defect engineering effect. AFM images shown in Figure 3 again confirm the morphology evolution with the increase of He doses. In the meantime, we probe the corresponding domain structures using piezoresponse force microscopy (PFM). The uniform out-of-plane (OOP) PFM contrast and the in-plane (IP) nanodomains shown in Figure 3a indicates the $M_c$ structure with a multi-domain state in the as-grown film, in consistent with previous work.[39,74-76] Both the implanted samples show the uniform OOP and IP contrasts (Figure 3b and 3c), indicating the formation of true T phase with a single domain state in agreement with our XRD and RSM data. Therefore, the multi-domain to single domain structure evolution further verifies the $M_c$-T phase transition (this observation is distinct from the discovery of multi- to a single-domain state change in $(Ba,Ca)(Zr,Ti)O_3$ under electric field that shows the ferroelectric rotation process without any structural change[77]). It is worth to note that although temperature-driven true T phase $BiFeO_3$ has been reported,[70,71] the domain structure has not been captured, inhibited by the unstable PFM scans at high temperature. Here we first time obtain the PFM images that reveal the single domain state in true T phase $BiFeO_3$.

To explore the reversibility of phase transition and domain evolution induced by He implantation, high temperature annealing of the $5 \times 10^{14}$ He/$cm^2$ dosed $BiFeO_3$ film was carried out at 550 °C under 1 atm oxygen atmosphere for 1 hour. A combination of re-emergence of dark-contrast R phase (Figure 4a) and nanoscale stripe domains (Figure 4c) indicates the reproduction of mixed R and Mc phases due to the release of He, which is confirmed with the (002), (103) and (113) RSMs shown in Figure 4d, e and f, respectively. Schematics shown in Figure 4g summarize the reversible control of R-Mc-T-super T phase transition in $BiFeO_3$ *via* He ion implantation and high temperature annealing.

In summary, we have demonstrated the ability to controllable defect-driven R-Mc-T-super T phase transition and multi-single domain evolution in $BiFeO_3$ thin film, enabling the creation of the single-domain-state, true super-tetragonal phase with enhanced tetragonality (c/a ~ 1.30). The reversible phase transition and domain evolution have been revealed by high temperature annealing to release the implanted point defects. Our findings open a door to achieve giant tetragonality (or c/a ratio) in ferroelectric thin films that could be extended to other ferroelectric material systems (such as $PbTiO_3$, $BaTiO_3$ and $HfO_2$) which might lead to strong polarization enhancement. Moreover, the discovery of controlled defect engineered domain evolution in this work, provides a future direction to create and manipulate exotic domain states (e.g., vortices, skyrmions) in ferroelectric and magnetic materials *via* this approach.

AUTHOR INFORMATION

**Corresponding Authors**

*E-mail: deyangchen@m.scnu.edu.cn (D. Chen)

*E-mail: xingsengao@scnu.edu.cn (X. Gao)

**Author Contributions**

D.C. conceived and designed the project. S.Z. proposed the ion implantation/irradiation approach. C.C and C.L. fabricated and characterized the thin films. C.W. and U.K. conducted the He implantation experiments. X.C., C.X., N.W. and Y.Z. carried out the STEM measurements. J.Z., Z.L. and P.Y. contributed to the XRD and RSM data interpretation. Z.F., M.Q., M.Z., X.L. and G.Z. analyzed the PFM data. X.G. and J.-M.L. discussed the TEM and AFM data. D.C. and C.C. wrote the paper with contributions and feedback from all authors. All authors discussed the results and commented on the manuscript.

**Notes**

The authors declare no competing financial interest.

**Acknowledgements**

This work was supported by the National Key Research and Development Program of China (No. 2016YFA0201002) and the National Natural Science Foundation of China (Grant Nos. U1832104 and 11704130). Authors also acknowledges the financial support of the Natural Science Foundation of Guangdong Province (No. 2017A30310169). Y. Z. thanks the financial support from the Research Grants Council of Hong Kong (Project No. 15305718) and the Hong Kong Polytechnic University grant (Project No. 1-ZE6G). N. W. thanks the financial support from the Research Grants Council of Hong Kong (Project No. C6021-14E). C.Wang thanks China Scholarship Council (No.201606750007) for financial supports. S. Z. acknowledges financial support from the German Research Foundation (ZH 225/10-1). Ion irradiation was done at the Ion Beam Center (IBC) at Helmholtz-Zentrum Dresden-Rossendorf. X. Gao and

X. Lu acknowledge the Project for Guangdong Province Universities and Colleges Pearl River Scholar Funded Scheme 2014 and 2016, respectively.

## REFERENCES

(1) Lee, C.-H.; Orloff, N. D.; Birol, T.; Zhu, Y.; Goian, V.; Rocas, E.; Haislmaier, R.; Vlahos, E.; Mundy, J. A.; Kourkoutis, L. F.; Nie, Y.; Biegalski, M. D.; Zhang, J.; Bernhagen, M.; Benedek, N. A.; Kim, Y.; Brock, J. D.; Uecker, R.; Xi, X. X.; Gopalan, V.; Nuzhnyy, D.; Kamba, S.; Muller, D. A.; Takeuchi, I.; Booth, J. C.; Fennie, C. J.; Schlom, D. G. *Nature* **2013**, *502*, 532-536.

(2) Wang, Z.; Guo, H. W.; Shao, S.; Saghayezhian, M.; Li, J.; Fittipaldi, R.; Vecchione, A.; Siwakoti, P.; Zhu, Y. M.; Zhang, J. D.; Plummer, E. W. *Proc. Natl. Acad. Sci. U. S. A.* **2018**, *115*, 9485-9490.

(3) Kalinin, S. V.; Spaldin, N. A. *Science* **2013**, *341*, 858-859.

(4) Tuller, H. L.; Bishop, S. R. *Ann. Rev. Mater. Res.* **2011**, *41*, 369-398.

(5) Huang, Z.; Ariando; Wang, X. R.; Rusydi, A.; Chen, J. S.; Yang, H.; Venkatesan, T. *Adv. Mater.* **2018**, *30*, 1802439.

(6) Saremi, S.; Gao, R.; Dasgupta, A.; Martin, L. W. *Am. Ceram. Soc. Bull.* **2018**, *97*, 16-23.

(7) Boschker, H.; Mannhart, J. *Annu. Rev. Conden. Matter Phys* **2017**, *8*, 145-164.

(8) Yadav, A. K.; Nelson, C. T.; Hsu, S. L.; Hong, Z.; Clarkson, J. D.; Schlepuetz, C. M.; Damodaran, A. R.; Shafer, P.; Arenholz, E.; Dedon, L. R.; Chen, D.; Vishwanath, A.; Minor, A. M.; Chen, L. Q.; Scott, J. F.; Martin, L. W.; Ramesh, R. *Nature* **2016**, *530*, 198-201.

(9) Gopalan, V.; Dierolf, V.; Scrymgeour, D. A. *Ann. Rev. Mater. Res.* **2007**, *37*, 449-489.

(10) Seidel, J.; Vasudevan, R. K.; Valanoor, N. *Adv. Electron. Mater.* **2016**, *2*, 1500292.

(11) Li, L. Z.; Zhang, Y.; Xie, L.; Jokisaari, J. R.; Beekman, C.; Yang, J. C.; Chu, Y. H.; Christen, H. M.; Pan, X. Q. *Nano Lett.* **2017**, *17*, 3556-3562.

(12) Kimmel, A. V.; Weaver, P. M.; Cain, M. G.; Sushko, P. V. *Phys. Rev. Lett.* **2012**, *109*, 117601.

(13) Zhang, L.; Chen, J.; Fan, L.; Diéguez, O.; Cao, J.; Pan, Z.; Wang, Y.; Wang, J.; Kim, M.; Deng, S.; Wang, J.; Wang, H.; Deng, J.; Yu, R.; Scott, J. F.; Xing, X. *Science* **2018**, *361*, 494-497.

(14) Xie, L.; Li, L. Z.; Heikes, C. A.; Zhang, Y.; Hong, Z. J.; Gao, P.; Nelson, C. T.; Xue, F.; Kioupakis, E.; Chen, L. Q.; Schlom, D. G.; Wang, P.; Pan, X. Q. *Adv. Mater.* **2017**, *29*, 1701475.

(15) Liu, H.; Yang, P.; Yao, K.; Ong, K. P.; Wu, P.; Wang, J. *Adv. Funct. Mater.* **2012**, *22*, 937-942.

(16) Sando, D.; Young, T.; Bulanadi, R.; Cheng, X.; Zhou, Y. Y.; Weyland, M.; Munroe, P.; Nagarajan, V. *Jpn. J. Appl. Phys.* **2018**, *57*, 0902B2.

(17) Li, L. Z.; Cheng, X. X.; Jokisaari, J. R.; Gao, P.; Britson, J.; Adamo, C.; Heikes, C.; Schlom, D. G.; Chen, L. Q.; Pan, X. Q. *Phys. Rev. Lett.* **2018**, *120*, 137602.

(18) Li, L. Z.; Jokisaari, J. R.; Zhang, Y.; Cheng, X. X.; Yan, X. X.; Heikes, C.; Lin, Q. Y.; Gadre, C.; Schlom, D. G.; Chen, L. Q.; Pan, X. Q. *Adv. Mater.* **2018**, *30,* 1802737.

(19) Rojac, T.; Bencan, A.; Drazic, G.; Sakamoto, N.; Ursic, H.; Jancar, B.; Tavcar, G.; Makarovic, M.; Walker, J.; Malic, B.; Damjanovic, D. *Nat. Mater.* **2017**, *16*, 322-327.

(20) Kim, Y.-M.; Morozovska, A.; Eliseev, E.; Oxley, M. P.; Mishra, R.; Selbach, S. M.; Grande, T.; Pantelides, S. T.; Kalinin, S. V.; Borisevich, A. Y. *Nat. Mater.* **2014**, *13*, 1019-1025.

(21) Damodaran, A. R.; Breckenfeld, E.; Chen, Z. H.; Lee, S.; Martin, L. W. *Adv. Mater.* **2014**, *26*, 6341-6347.

(22) Kalinin, S. V.; Jesse, S.; Rodriguez, B. J.; Chu, Y. H.; Ramesh, R.; Eliseev, E. A.; Morozovska, A. N. *Phys. Rev. Lett.* **2008**, *100*, 155703.

(23) Chen, D.; Chen, Z.; He, Q.; Clarkson, J. D.; Serrao, C. R.; Yadav, A. K.; Nowakowski, M. E.; Fan, Z.; You, L.; Gao, X.; Zeng, D.; Chen, L.; Borisevich, A. Y.; Salahuddin, S.; Liu, J.-M.; Bokor, J. *Nano Lett.* **2017**, *17*, 486-493.

(24) Dearnaley, G. *Nature* **1975**, *256*, 701.

(25) Shinada, T.; Okamoto, S.; Kobayashi, T.; Ohdomari, I. *Nature* **2005**, *437*, 1128-1131.

(26) Yoon, K.; Rahnamoun, A.; Swett, J. L.; Iberi, V.; Cullen, D. A.; Vlassiouk, I. V.; Belianinov, A.; Jesse, S.; Sang, X.; Ovchinnikova, O. S.; Rondinone, A. J.; Unocic, R. R.; van Duin, A. C. T. *ACS Nano* **2016**, *10*, 8376-8384.

(27) Wang, J.; Neaton, J. B.; Zheng, H.; Nagarajan, V.; Ogale, S. B.; Liu, B.; Viehland, D.; Vaithyanathan, V.; Schlom, D. G.; Waghmare, U. V.; Spaldin, N. A.; Rabe, K. M.; Wuttig, M.; Ramesh, R. *Science* **2003**, *299*, 1719-1722.

(28) Catalan, G.; Scott, J. F. *Adv. Mater.* **2009**, *21*, 2463-2485.

(29) Zeches, R. J.; Rossell, M. D.; Zhang, J. X.; Hatt, A. J.; He, Q.; Yang, C.-H.; Kumar, A.; Wang, C. H.; Melville, A.; Adamo, C.; Sheng, G.; Chu, Y.-H.; Ihlefeld, J. F.; Erni, R.; Ederer, C.; Gopalan, V.; Chen, L. Q.; Schlom, D. G.; Spaldin, N. A.; Martin, L. W.; Ramesh, R. *Science* **2009**, *326*, 977-980.

(30) Yang, J. C.; He, Q.; Suresha, S. J.; Kuo, C. Y.; Peng, C. Y.; Haislmaier, R. C.; Motyka, M. A.; Sheng, G.; Adamo, C.; Lin, H. J.; Hu, Z.; Chang, L.; Tjeng, L. H.; Arenholz, E.; Podraza, N. J.; Bernhagen, M.; Uecker, R.; Schlom, D. G.; Gopalan, V.; Chen, L. Q.; Chen, C. T.; Ramesh, R.; Chu, Y. H. *Phys. Rev. Lett.* **2012**, *109*, 247606.

(31) Christen, H. M.; Nam, J. H.; Kim, H. S.; Hatt, A. J.; Spaldin, N. A. *Phys. Rev. B* **2011**, *83*, 144107.

(32) Chen, Z.; Prosandeev, S.; Luo, Z. L.; Ren, W.; Qi, Y.; Huang, C. W.; You, L.; Gao, C.; Kornev, I. A.; Wu, T.; Wang, J.; Yang, P.; Sritharan, T.; Bellaiche, L.; Chen, L. *Phys. Rev. B* **2011**, *84*, 094116.

(33) Bea, H.; Dupe, B.; Fusil, S.; Mattana, R.; Jacquet, E.; Warot-Fonrose, B.; Wilhelm, F.; Rogalev, A.; Petit, S.; Cros, V.; Anane, A.; Petroff, F.; Bouzehouane, K.; Geneste, G.; Dkhil, B.; Lisenkov, S.; Ponomareva, I.; Bellaiche, L.; Bibes, M.; Barthelemy, A. *Phys. Rev. Lett.* **2009**, *102*, 217603.

(34) Chen, D.; Nelson, C. T.; Zhu, X.; Serrao, C. R.; Clarkson, J. D.; Wang, Z.; Gao, Y.; Hsu, S.-L.; Dedon, L. R.; Chen, Z.; Yi, D.; Liu, H.-J.; Zeng, D.; Chu, Y.-H.; Liu, J.; Schlom, D. G.; Ramesh, R. *Nano Lett.* **2017**, *17*, 5823-5829.

(35) Damodaran, A. R.; Liang, C.-W.; He, Q.; Peng, C.-Y.; Chang, L.; Chu, Y.-H.; Martin, L. W. *Adv. Mater.* **2011**, *23*, 3170-3175.

(36) Belik, A. A.; Yusa, H.; Hirao, N.; Ohishi, Y.; Takayama-Muromachi, E. *Chem. Mater.* **2009**, *21*, 3400-3405.
(37) Knee, C. S.; Tucker, M. G.; Manuel, P.; Cai, S.; Bielecki, J.; Börjesson, L.; Eriksson, S. G. *Chem. Mater.* **2014**, *26*, 1180-1186.
(38) Sando, D.; Xu, B.; Bellaiche, L.; Nagarajan, V. *Appl. Phys. Rev.* **2016**, *3*, 011106.
(39) Chen, Z. H.; Luo, Z. L.; Huang, C. W.; Qi, Y. J.; Yang, P.; You, L.; Hu, C. S.; Wu, T.; Wang, J. L.; Gao, C.; Sritharan, T.; Chen, L. *Adv. Funct. Mater.* **2011**, *21*, 133-138.
(40) Rusakov, D. A.; Abakumov, A. M.; Yamaura, K.; Belik, A. A.; Van Tendeloo, G.; Takayama-Muromachi, E. *Chem. Mater.* **2011**, *23*, 285-292.
(41) Kan, D.; Pálová, L.; Anbusathaiah, V.; Cheng, C. J.; Fujino, S.; Nagarajan, V.; Rabe, K. M.; Takeuchi, I. *Adv. Funct. Mater.* **2010**, *20*, 1108-1115.
(42) Yang, C.-H.; Kan, D.; Takeuchi, I.; Nagarajan, V.; Seidel, J. *Phys. Chem. Chem. Phys.* **2012**, *14*, 15953-15962.
(43) Zhang, J. X.; He, Q.; Trassin, M.; Luo, W.; Yi, D.; Rossell, M. D.; Yu, P.; You, L.; Wang, C. H.; Kuo, C. Y.; Heron, J. T.; Hu, Z.; Zeches, R. J.; Lin, H. J.; Tanaka, A.; Chen, C. T.; Tjeng, L. H.; Chu, Y. H.; Ramesh, R. *Phys. Rev. Lett.* **2011**, *107*, 147602.
(44) Chu, K.; Jang, B.-K.; Sung, J. H.; Shin, Y. A.; Lee, E.-S.; Song, K.; Lee, J. H.; Woo, C.-S.; Kim, S. J.; Choi, S.-Y.; Koo, T. Y.; Kim, Y.-H.; Oh, S.-H.; Jo, M.-H.; Yang, C.-H. *Nat. Nanotechnol.* **2015**, *10*, 972.
(45) Fujino, S.; Murakami, M.; Anbusathaiah, V.; Lim, S.-H.; Nagarajan, V.; Fennie, C. J.; Wuttig, M.; Salamanca-Riba, L.; Takeuchi, I. *Appl. Phys. Lett.* **2008**, *92*, 202904.
(46) He, Q.; Chu, Y. H.; Heron, J. T.; Yang, S. Y.; Liang, W. I.; Kuo, C. Y.; Lin, H. J.; Yu, P.; Liang, C. W.; Zeches, R. J.; Kuo, W. C.; Juang, J. Y.; Chen, C. T.; Arenholz, E.; Scholl, A.; Ramesh, R. *Nat. Commun.* **2011**, *2*, 225.
(47) Infante, I. C.; Juraszek, J.; Fusil, S.; Dupé, B.; Gemeiner, P.; Diéguez, O.; Pailloux, F.; Jouen, S.; Jacquet, E.; Geneste, G.; Pacaud, J.; Íñiguez, J.; Bellaiche, L.; Barthélémy, A.; Dkhil, B.; Bibes, M. *Phys. Rev. Lett.* **2011**, *107*, 237601.
(48) Seidel, J.; Trassin, M.; Zhang, Y.; Maksymovych, P.; Uhlig, T.; Milde, P.; Köhler, D.; Baddorf, A. P.; Kalinin, S. V.; Eng, L. M.; Pan, X.; Ramesh, R. *Adv. Mater.* **2014**, *26*, 4376-4380.
(49) Jang, B.-K.; Lee, J. H.; Chu, K.; Sharma, P.; Kim, G.-Y.; Ko, K.-T.; Kim, K.-E.; Kim, Y.-J.; Kang, K.; Jang, H.-B.; Jang, H.; Jung, M. H.; Song, K.; Koo, T. Y.; Choi, S.-Y.; Seidel, J.; Jeong, Y. H.; Ohldag, H.; Lee, J.-S.; Yang, C.-H. *Nat. Phys.* **2016**, *13*, 189.
(50) Kim, K.-E.; Jeong, S.; Chu, K.; Lee, J. H.; Kim, G.-Y.; Xue, F.; Koo, T. Y.; Chen, L.-Q.; Choi, S.-Y.; Ramesh, R.; Yang, C.-H. *Nat. Commun.* **2018**, *9*, 403.
(51) You, L.; Zheng, F.; Fang, L.; Zhou, Y.; Tan, L. Z.; Zhang, Z. Y.; Ma, G. H.; Schmidt, D.; Rusydi, A.; Wang, L.; Chang, L.; Rappe, A. M.; Wang, J. L. *Sci. Adv.* **2018**, *4*, eaat3438.
(52) Ederer, C.; Spaldin, N. A. *Phys. Rev. Lett.* **2005**, *95*, 257601.
(53) Hatt, A. J.; Spaldin, N. A.; Ederer, C. *Phys. Rev. B* **2010**, *81*, 054109.
(54) Qi, T.; Grinberg, I.; Rappe, A. M. *Phys. Rev. B* **2009**, *79*, 094114.
(55) Qi, T.; Grinberg, I.; Rappe, A. M. *Phys. Rev. B* **2010**, *82*, 134113.
(56) Kitanaka, Y.; Ogino, M.; Noguchi, Y.; Miyayama, M.; Hoshikawa, A.; Ishigaki, T. *Jpn. J. Appl. Phys.* **2018**, *57*, 11UD05.

(57) Fan, Z.; Xiao, J.; Liu, H.; Yang, P.; Ke, Q.; Ji, W.; Yao, K.; Ong, K. P.; Zeng, K.; Wang, J. *ACS Appl. Mater. Inter.* **2015**, *7*, 2648-2653.
(58) Jia, C.-L.; Nagarajan, V.; He, J.-Q.; Houben, L.; Zhao, T.; Ramesh, R.; Urban, K.; Waser, R. *Nat. Mater.* **2006**, *6*, 64.
(59) Wei, Y.; Nukala, P.; Salverda, M.; Matzen, S.; Zhao, H. J.; Momand, J.; Everhardt, A. S.; Agnus, G.; Blake, G. R.; Lecoeur, P.; Kooi, B. J.; Íñiguez, J.; Dkhil, B.; Noheda, B. *Nat. Mater.* **2018**, *17*, 1095-1100.
(60) Takamura, Y.; Chopdekar, R. V.; Scholl, A.; Doran, A.; Liddle, J. A.; Harteneck, B.; Suzuki, Y. *Nano Lett.* **2006**, *6*, 1287-1291.
(61) Guo, H. W.; Dong, S.; Rack, P. D.; Budai, J. D.; Beekman, C.; Gai, Z.; Siemons, W.; Gonzalez, C. M.; Timilsina, R.; Wong, A. T.; Herklotz, A.; Snijders, P. C.; Dagotto, E.; Ward, T. Z. *Phys. Rev. Lett.* **2015**, *114*, 256801.
(62) Herklotz, A.; Rus, S. F.; Ward, T. Z. *Nano Lett.* **2016**, *16*, 1782-1786.
(63) Wang, C.; Chen, C.; Chang, C.-H.; Tsai, H.-S.; Pandey, P.; Xu, C.; Böttger, R.; Chen, D.; Zeng, Y.-J.; Gao, X.; Helm, M.; Zhou, S. *ACS Appl. Mater. Inter.* **2018**, *10*, 27472-27476.
(64) Herklotz, A.; Gai, Z.; Sharma, Y.; Huon, A.; Rus, S. F.; Sun, L.; Shen, J.; Rack, P. D.; Ward, T. Z. *Adv. Sci.* **2018**, *5*, 1800356.
(65) Saremi, S.; Xu, R.; Dedon, L. R.; Mundy, J. A.; Hsu, S.-L.; Chen, Z.; Damodaran, A. R.; Chapman, S. P.; Evans, J. T.; Martin, L. W. *Adv. Mater.* **2016**, *28*, 10750–10756.
(66) Saremi, S.; Xu, R.; Dedon, L. R.; Gao, R.; Ghosh, A.; Dasgupta, A.; Martin, L. W. *Adv. Mater. Interfaces* **2018**, *5*, 1700991.
(67) Saremi, S.; Xu, R.; Allen, F. I.; Maher, J.; Agar, J. C.; Gao, R.; Hosemann, P.; Martin, L. W. *Phys. Rev. Mater.* **2018**, *2*, 084414.
(68) Pailloux, F.; Couillard, M.; Fusil, S.; Bruno, F.; Saidi, W.; Garcia, V.; Carrétéro, C.; Jacquet, E.; Bibes, M.; Barthélémy, A.; Botton, G. A.; Pacaud, J. *Phys. Rev. B* **2014**, *89*, 104106.
(69) Liu, H.-J.; Du, Y.-H.; Gao, P.; Huang, Y.-C.; Chen, H.-W.; Chen, Y.-C.; Liu, H.-L.; He, Q.; Ikuhara, Y.; Chu, Y.-H. *APL Mater.* **2015**, *3*, 116104.
(70) Liu, H.-J.; Chen, H.-J.; Liang, W.-I.; Liang, C.-W.; Lee, H.-Y.; Lin, S.-J.; Chu, Y.-H. *J. Appl. Phys.* **2012**, *112*, 052002.
(71) Beekman, C.; Siemons, W.; Ward, T. Z.; Chi, M.; Howe, J.; Biegalski, M. D.; Balke, N.; Maksymovych, P.; Farrar, A. K.; Romero, J. B.; Gao, P.; Pan, X. Q.; Tenne, D. A.; Christen, H. M. *Adv. Mater.* **2013**, *25*, 5561-5567.
(72) Zhang, J.; Ke, X.; Gou, G.; Seidel, J.; Xiang, B.; Yu, P.; Liang, W.-I.; Minor, A. M.; Chu, Y.-h.; Van Tendeloo, G.; Ren, X.; Ramesh, R. *Nat. Commun.* **2013**, *4*, 2768.
(73) Zhang, J. X.; Zeches, R. J.; He, Q.; Chu, Y. H.; Ramesh, R. *Nanoscale* **2012**, *4*, 6196-6204.
(74) Mazumdar, D.; Shelke, V.; Iliev, M.; Jesse, S.; Kumar, A.; Kalinin, S. V.; Baddorf, A. P.; Gupta, A. *Nano Lett.* **2010**, *10*, 2555-2561.
(75) You, L.; Chen, Z. H.; Zou, X.; Ding, H.; Chen, W. G.; Chen, L.; Yuan, G. L.; Wang, J. L. *ACS Nano* **2012**, *6*, 5388-5394.
(76) Trassin, M.; Luca, G. D.; Manz, S.; Fiebig, M. *Adv. Mater.* **2015**, *27*, 4871-4876.
(77) Zakhozheva, M.; Schmitt, A.; Acosta, M.; Jo, W.; Rödel, J.; Kleebe, H.-J. *Appl. Phys. Lett.* **2014**, *105*, 112904.

View Article Online
DOI: 10.1039/C9NR00932A

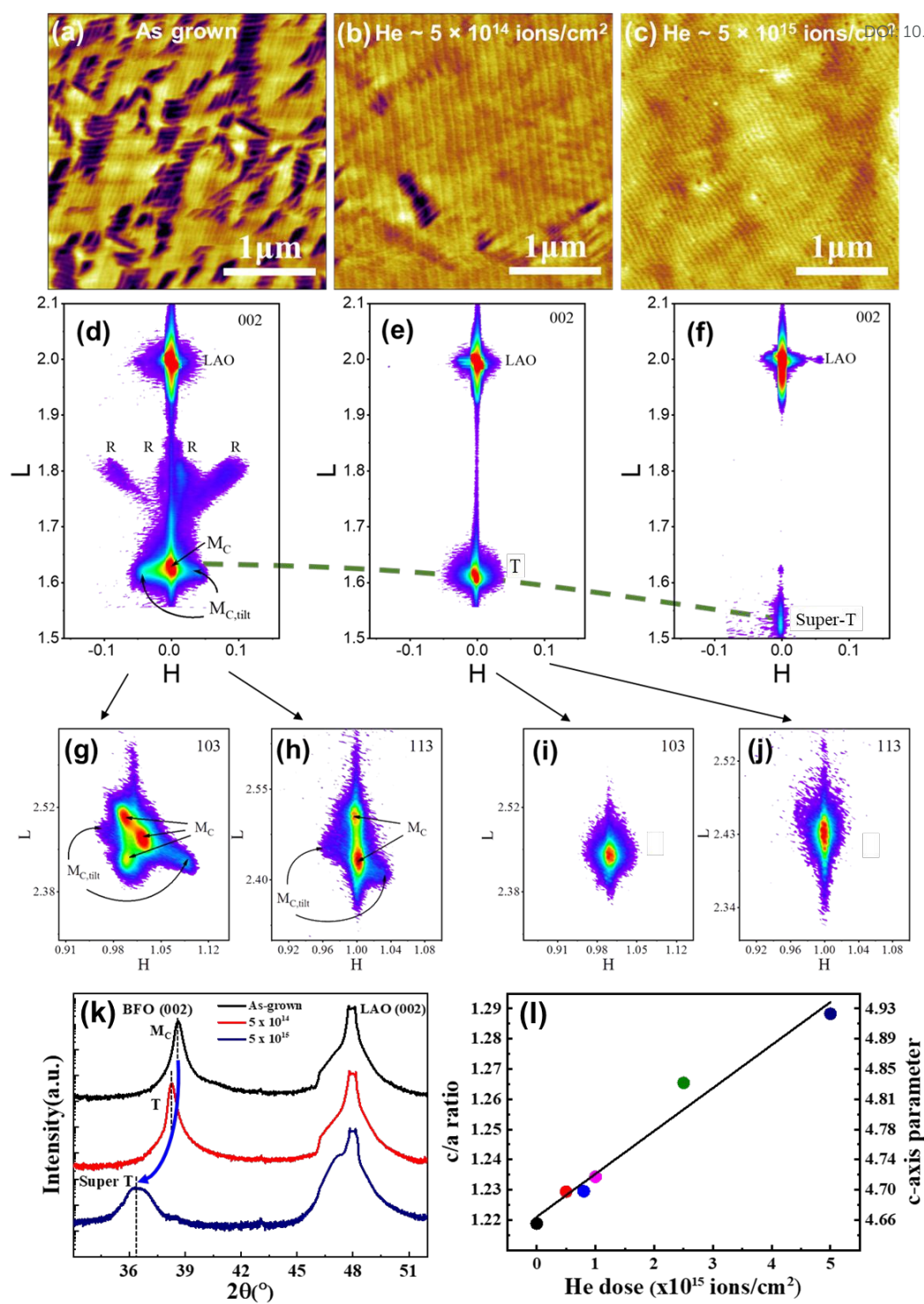

**Figure 1.** (a-c) Morphology evolution in mixed phase $BiFeO_3$ (BFO) with the increase of implanted He dosages. (d-f) RSM (002) reflections corresponding to (a-c) samples reveal the He implantation (defect engineering) induced phase transitions and elongated out-of-plane (c-axis). (g-h) Three-fold and two-fold splits along (103) and (113) RSMs confirm the tetragonal-like Mc phase in as-grown samples. (i-j) Single peak feature of (103) and (113) RSMs in the He implanted sample indicates the creation of true tetragonal phase. (k) Typical X-ray θ–2θ scans of as-grown and He implanted $BiFeO_3$ thin films on $LaAlO_3$ (LAO) substrates. (l) Dependence of c/a ratio and c-axis lattice parameter with various He dosages.

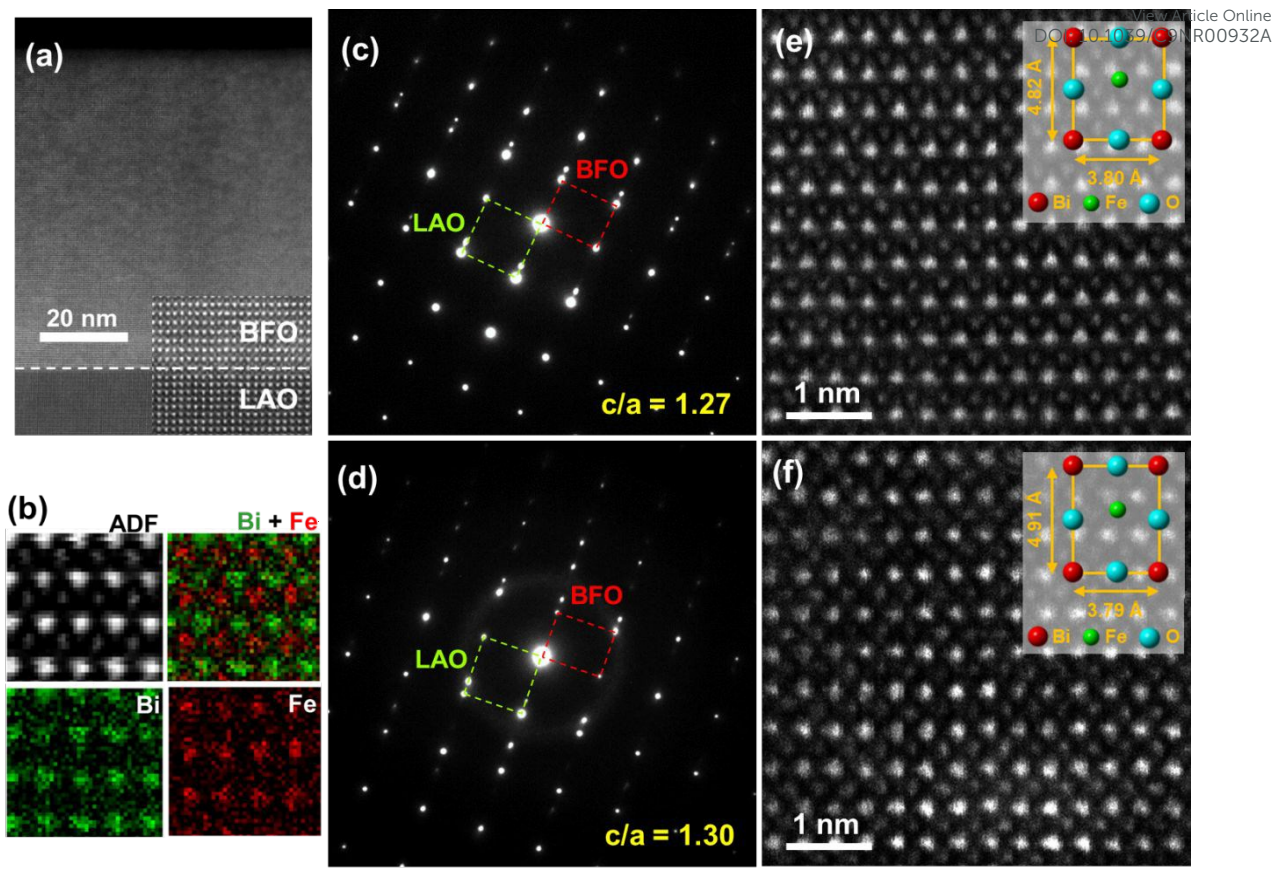

**Figure 2.** (a) Dark field cross-sectional TEM images of a He implanted 70-nm-thick BiFeO$_3$ thin film and BiFeO$_3$ and LaAlO$_3$ interface (insets in the bottom-right). (b) Angular dark field (ADF)-STEM image of BiFeO$_3$ and the corresponding EDS mapping. (c-d) Selected area electron diffraction (SAED) patterns of implanted BiFeO$_3$ with He doses of 2.5 × 10$^{15}$ and 5 × 10$^{15}$ ions/cm$^2$, respectively. (e-f) ADF-STEM atomic structure images correspond to the SAED samples in (c-d).

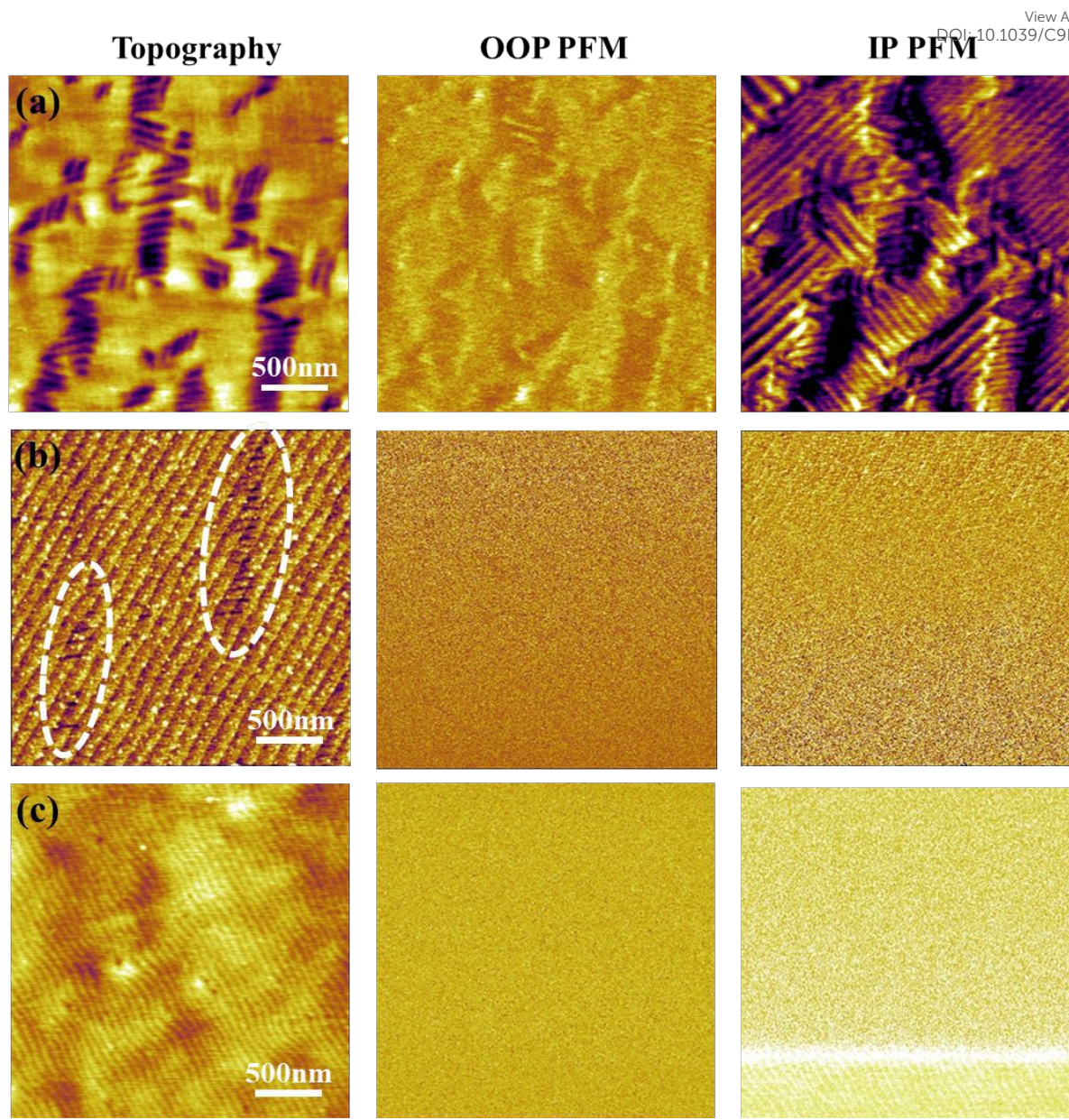

**Figure 3.** Topography, out-of-plane (OOP) and in-plane (IP) PFM images (from left to right) reveal both the morphology and domain structure evolution, with the increasing He implantation doses, in 70-nm-thick $BiFeO_3$ thin films: (a) Mixed phase morphology with IP PFM nanoscale stripe domains in the as-grown $BiFeO_3$ sample. (b-c) Morphologies and PFM images of implanted samples show phase transition and domain structure evolution through He implantation with doses of $5 \times 10^{14}$ and $5 \times 10^{15}$ ions/cm$^2$, respectively.

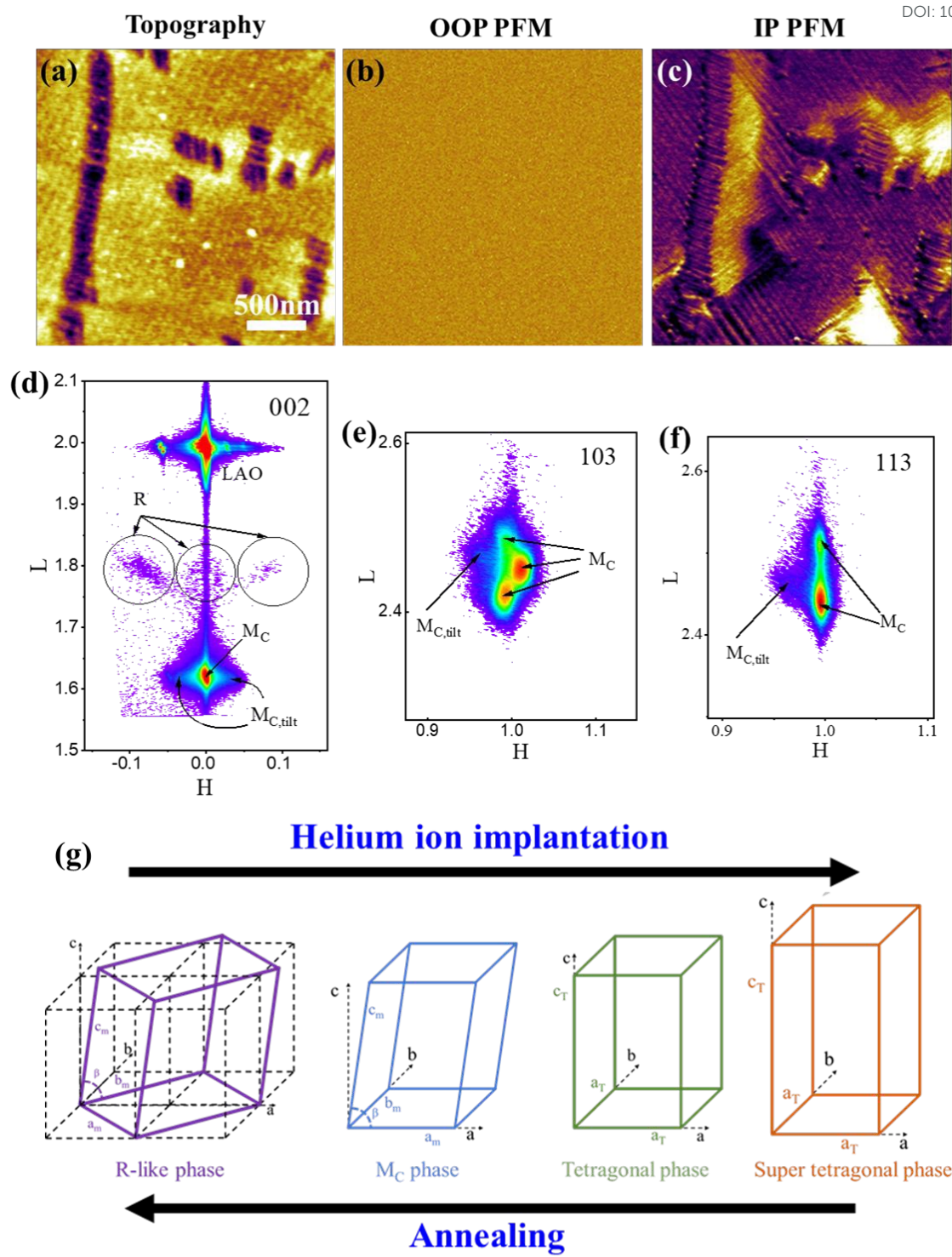

**Figure 4.** (a) Topography, (b) OOP PFM and (c) IP PFM images after annealing the implanted $BiFeO_3$ at 550 °C under 1 atm oxygen atmosphere for 1 hour. (d) (002), (e) (103) and (f) (113) RSM reflections of the annealing sample shown in (a-c). (g) Schematics of reversible control of R-Mc-T-super T phase transition in $BiFeO_3$ *via* He ion implantation and annealing that have been demonstrated in this work.

**Table of Contents**

Single-domain-state true super-tetragonal BiFeO$_3$ phase with the largest c/a ratio ~ 1.3 has been achieved through controllable defect engineering.

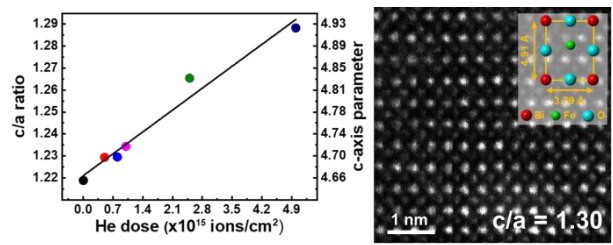